# Seebeck Effect at the Atomic Scale

Eui-Sup Lee,[1] Sanghee Cho,[2] Ho-Ki Lyeo,[2,*] and Yong-Hyun Kim[1,*]

[1]*Graduate School of Nanoscience and Technology, KAIST, Daejeon 305-701, Republic of Korea*

[2]*Korea Research Institute of Standards and Science, Daejeon 305-340, Republic of Korea*

*Corresponding Authors: yong.hyun.kim@kaist.ac.kr; hklyeo@kriss.re.kr

**Abstract**

The atomic variations of electronic wavefunctions at the surface and electron scattering near a defect have been detected unprecedentedly by tracing thermoelectric voltages given a temperature bias [Cho *et al*., Nature Mater. **12**, 913 (2013)]. Because thermoelectricity, or Seebeck effect, is associated with heat-induced electron diffusion, how the thermoelectric signal is related to the atomic-scale wavefunctions and what the role of the temperature is at such a length scale remain very unclear. Here we show that coherent electron and heat transport through a point-like contact produces an atomic Seebeck effect, which is described by mesoscopic Seebeck coefficient multiplied with an effective temperature drop at the interface. The mesoscopic Seebeck coefficient is approximately proportional to the logarithmic energy derivative of local density of states at the Fermi energy. We deduced that the effective temperature drop at the tip-sample junction could vary at a sub-angstrom scale depending on atom-to-atom interaction at the interface. A computer-based simulation method of thermoelectric images is proposed, and a point defect in graphene was identified by comparing experiment and the simulation of thermoelectric imaging.

The invention of the scanning tunneling microscope (STM) by Binnig and Rohrer [1,2] facilitated direct access to microscopic quantum mechanics [3]. This method provides *real-space* wavefunction images of a material surface by measuring the electrical tunneling currents across the vacuum gap. The microscopic imaging mechanism of atomically resolved wavefunctions in the tunneling microscopy is rather straightforward [4] because the tunneling current can be easily localized in space by controlling the vacuum gap.

Heat, a measure of entropy, is largely perceived to be diffusive and transported incoherently by charge carriers (electrons and holes) and lattice vibrations (or phonons) in a material. Heat transport is therefore considered a challenging means of the local imaging of a material and its electronic states [5,6]. Very recently, however, Cho *et al*. [7] reported a series of atomic wavefunction images of epitaxial graphene, obtained while performing local thermoelectric imaging with a heat-based scanning probe microscope [6,7]. These counter-intuitive heat-based *real-space* wavefunction images naturally generate one key question: how can one measure the atomic variation in the unit cell in a heat transport experiment? To answer this question, we must not only elucidate the imaging mechanism of the scanning thermoelectric microscope, but also re-evaluate the fundamental physics of thermoelectricity, or Seebeck effect, from conventional length scales to the atomic length scale.

In this Letter, we present a theory of scanning thermoelectric microscopy with atomic resolution based on the mesoscopic electron and heat transport characteristics. This theory, beginning with the macroscopic general transport equation and electrostatic equations, illustrates the feasibility and mechanisms in play when observing atomically varying features with thermoelectric measurements. Computer simulations of thermoelectric images were efficaciously used for identifying atomic-scale defects in graphene in conjunction with experimental results.

The scanning thermoelectric microscope [7], a modified ultra-high vacuum contact-mode atomic force microscope (AFM), operates with a conductive probe and with a sample at temperatures of $T_1$ and $T_2$, as schematically shown in Fig. 1. The temperature difference between the AFM tip and the sample induces a localized thermoelectric voltage that is measured with a high-impedance voltmeter. The local thermoelectric voltage measured in the AFM setup can be represented by

$$V(\mathbf{r}) = V_{\text{diff}} + S_{\text{coh}}(\mathbf{r})\Delta T_{\text{coh}}(\mathbf{r}), \qquad (1)$$

where $V_{\text{diff}}$ is a thermoelectric voltage drop in the diffusive transport region both at the tip and sample, and $S_{\text{coh}}(\mathbf{r})$ and $\Delta T_{\text{coh}}(\mathbf{r})$ are position-dependent Seebeck coefficient and effective temperature drop at the interface between the tip and sample, where electron and heat transport coherently. The coherent thermoelectric voltage $V_{\text{coh}}(\mathbf{r}) = S_{\text{coh}}(\mathbf{r})\Delta T_{\text{coh}}(\mathbf{r})$ is responsible for the atomic resolution observed in the scanning thermoelectric microscope [7], as we will discuss below.

When a temperature gradient $\nabla T$ is present in a macroscopic electro-conductive system, the transport of electrons or charged particles is subjective to an electrostatic field $\mathbf{E}$ and a driving force for particle diffusion under the gradient $\nabla T$ [8]. The electric current density $\mathbf{J}(\mathbf{r})$ at a local site $\mathbf{r}$ is expressed by the general transport equations as $\mathbf{J}(\mathbf{r}) = \sigma[\mathbf{E}(\mathbf{r}) - S(\mathbf{r})\nabla T(\mathbf{r})]$, where $\sigma$ is the electrical conductivity and $S(\mathbf{r})$ is the local Seebeck coefficient or thermopower [8]. At the *open-circuit* limit, $\mathbf{J}(\mathbf{r})=0$, as in the case of an ideal voltmeter, the charged particles then experience a balance between the electrostatic force and the thermopower force, represented by

$$\mathbf{E}(\mathbf{r}) = S(\mathbf{r})\nabla T(\mathbf{r}). \qquad (2)$$

The temperature profile $T(\mathbf{r})$ is primarily determined by the thermal transport properties of the system, such as the thermal conductivities of the constituent materials and the interfacial thermal conductance between the materials [5,9]. Although this force balance equation is derived for macroscopic diffusive systems, we postulate that Eq. (2) holds for microscopic systems with no net flow of electric current, in particular, across the interface between the tip and the sample at different temperatures (Fig. 1).

When no external electric field is applied, $\mathbf{E}(\mathbf{r})$ denotes only the built-in electric field resulting from the thermal diffusion-induced charge distribution $q^{th}(\mathbf{r})$. Gauss's law can be applied to the built-in electric field and the charge density; i.e., $\nabla \cdot \mathbf{E}(\mathbf{r}) = q^{th}(\mathbf{r})$. Then, Eq. (2) leads to

$$\nabla \cdot (S(\mathbf{r})\nabla T(\mathbf{r})) = q^{th}(\mathbf{r}). \qquad (3)$$

From this equation, the distribution of the heat-induced charge density, $q^{th}(\mathbf{r})$, can be accurately traced back from the information of the local Seebeck coefficient $S(\mathbf{r})$ and temperature profile $T(\mathbf{r})$.

When an AFM tip is used for generating and point-probing thermoelectric voltage as shown in Fig. 1, the local thermoelectric voltage is equivalent to the 'Hartree-type' electrostatic potential $V(\mathbf{r}) = \int q^{th}(\mathbf{r};\mathbf{r}')/|\mathbf{r}'-\mathbf{r}| d^3\mathbf{r}'$, where $\mathbf{r}$ is the position of the voltage probe and the integration over $\mathbf{r}'$ is for the entire volume. Then, using Eq. (3) and the fact that $\nabla T = 0$ at infinity, the local thermoelectric voltage is expressed (after integration by parts) as

$$V(\mathbf{r}) = \int S(\mathbf{r};\mathbf{r}')\nabla T(\mathbf{r};\mathbf{r}') \cdot \frac{\mathbf{r}'-\mathbf{r}}{|\mathbf{r}'-\mathbf{r}|^3} d^3\mathbf{r}'. \qquad (4)$$

The local thermoelectric voltage $V(\mathbf{r})$ is expressed in the form of a volume integral – not a line integral – of the local Seebeck coefficient $S(\mathbf{r};\mathbf{r}')$ convoluted by a radially weighted temperature gradient $\nabla T(\mathbf{r};\mathbf{r}')$ by a factor of $1/r^2$, where $r$ is a distance as measured from the point voltage probe. Because Eq. (4) is an exact expression that does not involve any approximation, it can be generally applicable for other thermoelectric systems including non-contact STM setups [10,11,12].

The volume integral in Eq. (4) can be split into the diffusive and coherent transport regions. In the diffusive transport region, the Seebeck coefficient and temperature profile are determined only by material properties such as electrical and thermal conductivities. From the Mott formula [13,14], the diffusive Seebeck coefficient is $S = -\frac{(k_B\pi)^2 T}{3e}\frac{\partial}{\partial E}\left(\ln[\sigma(E)]\right)_{E_F}$, where $k_B$ is the Boltzmann constant, $T$ is the absolute temperature, $e$ is the electron charge, $\sigma(E)$ is the energy-dependent electrical conductivity, and $E_F$ is the Fermi energy. The temperature profile $T(\mathbf{r};\mathbf{r}')$ will mostly vary slowly, governed by the phonon mean-free-path [9]. Then, the volume integral in Eq. (4) produces an almost constant thermoelectric voltage, termed as $V_{\text{diff}}$, by assuming a spherical temperature profile from the point thermal contact.

In the coherent transport region at the tip-sample interface, the transport of electrons and phonons across the junction can be accounted for by each transmission probability and electrical and thermal conductance quanta [15,16]. The coherent Seebeck coefficient $S_{\text{coh}}(\mathbf{r})$ is dependent only on the tip position $\mathbf{r}$ (independent of the internal coordinate $\mathbf{r}'$) and expressed in terms of the electron transmission probability $\tau(E,\mathbf{r})$ of the tip-sample junction as $S_{\text{coh}}(\mathbf{r}) = -\frac{(k_B\pi)^2 T}{3e}\frac{\partial}{\partial E}\left(\ln[\tau(E,\mathbf{r})]\right)_{E_F}$ from the Landauer formula [17-21]. The local coherent thermoelectric voltage is

$$V_{\text{coh}}(\mathbf{r}) = S_{\text{coh}}(\mathbf{r})\int \nabla T(\mathbf{r};\mathbf{r}') \cdot \frac{\mathbf{r}'-\mathbf{r}}{|\mathbf{r}'-\mathbf{r}|^3} d^3\mathbf{r}'$$

$$\equiv S_{\text{coh}}(\mathbf{r})\Delta T_{\text{coh}}(\mathbf{r}), \qquad (5)$$

where the volume integral of the weighted temperature profile is set to $\Delta T_{\text{coh}}(\mathbf{r})$, which is sensitively dependent on local geometry and near-probe temperature profile due to the $1/r^2$ weighting factor. Longer scale average of the effective temperature drop $\Delta T_{\text{coh}}(\mathbf{r})$ could correspond to the interfacial temperature drop that is known to exist at the thermal boundary between two different materials [9,16,22]. The interfacial temperature drop is often subjective to vibrational spectra [9,23,24] and interaction strength [25,26] of the involved materials.

With the diffusive thermoelectric voltage $V_{diff}$ and the coherent thermoelectric voltage in Eq. (5), we settled back to Eq. (1) for the total local thermoelectric voltage. Equation (5) is a rigorous expression of Seebeck effect at the atomic scale, advanced from the electron tunneling-based formulation by Stovneng and Lipavsky [11]. This atomic Seebeck effect explains how thermopower profiling works for semiconductor $p$-$n$ junctions [6] and how local thermoelectric imaging works seamlessly from micrometer to sub-angstrom scales [7].

At the weak-coupling limit, the electron transmission probability $\tau(E,\mathbf{r})$ can be approximated [4,11,15] as $\tau(E,\mathbf{r}) \propto N_e^{tip}(E) N_e^{sample}(E,\mathbf{r})$, where $N_e(E,\mathbf{r})$ is the local electronic density of states. From the Landauer formula [17-21], the coherent Seebeck coefficient $S_{coh}(\mathbf{r})$ can be expressed as $S_{coh}(\mathbf{r}) = S^{tip} + S^{sample}(\mathbf{r})$, owing to the logarithmic function. The Seebeck coefficient of sample, $S^{sample}(\mathbf{r})$, can also be written [17] as,

$$S^{sample}(\mathbf{r}) = -\frac{1}{eT} \frac{\int N_e^{sample}(E,\mathbf{r})(E-E_F)\left(-\frac{\partial f}{\partial E}\right)dE}{\int N_e^{sample}(E,\mathbf{r})\left(-\frac{\partial f}{\partial E}\right)dE}, \qquad (6)$$

where $f$ is the Fermi-Dirac distribution function at temperature $T$. At 300K, the $(-\partial f/\partial E)$ factor acts as an integration window of ~0.1 eV near the Fermi energy $E_F$. The sample Seebeck coefficient can be either positive or negative depending on the asymmetry or the energy derivative of $N_e^{sample}(E,\mathbf{r})$ at the Fermi energy [7].

Because we can obtain local density of states $N_e^{sample}(E,\mathbf{r})$ from first-principles calculations for a material surface, Eq. (1) should serve as a foundation of thermoelectric image simulation if we know the effective temperature drop $\Delta T_{coh}(\mathbf{r})$ at the tip-sample junction. Unfortunately, the information of $\Delta T_{coh}(\mathbf{r})$ is unknown from either experiment or simulation. We therefore deduced $\Delta T_{coh}(\mathbf{r})$ by inverting Eq. (1) with experimental $V^{expt}(\mathbf{r})$ and theoretical $S^{\text{sample}}(\mathbf{r})$ for a reference part of the sample, as shown in Fig. 2. With these results and first-principles calculations, we could reproduce experimental images and interpret the role of thermodynamic variables more clearly as discussed below.

Figure 2a and 2b display the measured thermoelectric voltage $V^{\text{expt}}(\mathbf{r})$ for a defect-free region in bilayer graphene on SiC (see Ref. [27] for experimental details). Figure 2c displays the theoretical Seebeck coefficient $S^{\text{sample}}(\mathbf{r})$ calculated from Eq. (6) with first-principles wavefunctions for *n*-doped free-standing graphene (see Ref. [28] for calculation details). Because the diamond-coated AFM tip and the graphene sample can interact through the van der Waals (vdW) interaction, the Seebeck coefficient $S^{\text{sample}}(\mathbf{r})$ was evaluated at the vdW equilibrium height as shown in Fig. S1 [28], which was calculated with the Lennard-Jones potential [29,30]. The experimental and theoretical thermoelectric images in Fig. 2b and 2c show a reasonable degree of correspondence. In particular, the center void of the carbon hexagon exhibits more negative signals in both the experimental and theoretical results than the carbon atom sites. This is characteristically different from STM, which usually picks up more current at charge-abundant atomic sites. By contrast, it is clear from line profiling in Fig. 2d that the Seebeck coefficient $S^{\text{sample}}(\mathbf{r})$ alone cannot reproduce the atomic corrugation observed in the thermoelectric voltage $V^{\text{expt}}(\mathbf{r})$.

To identify the role of temperature at the length scale of the coherent transport, we inverted Eq. (1) and deduced $\Delta T_{\text{coh}}(\mathbf{r})$ with $V^{\text{expt}}(\mathbf{r})$, $S^{\text{sample}}(\mathbf{r})$, $V_{\text{diff}} = -1.5$ mV, and $S^{\text{tip}} = 100$ µV/K [28,31]. We have found that there exists an almost linear correlation between the deduced $\Delta T_{\text{coh}}(\mathbf{r})$ and the vdW energy $E^{\text{vdW}}(\mathbf{r})$, as shown in Fig. 2e. The correlation indeed makes sense because inter-atomic thermal coupling that determines the temperature drop at the tip-sample interface may sensitively depend on the vdW interaction. When the thermal coupling or vdW interaction is relatively weak on top of carbon atom site, a larger interfacial temperature drop is expected than on the void. On the other hand, when the coupling is relatively strong, inter-atomic thermal conductance at the interface increases and thus temperature drop decreases. As a result, the deduced $\Delta T_{coh}(\mathbf{r})$ necessarily exhibits an atomic variation at a sub-angstrom scale from the carbon atom site to the hexagonal void, as shown in Fig. 2(d). Therefore, we propose that the atomic corrugation in the local thermoelectric voltage originates from the atomic variation in the coherent electron transport for the Seebeck coefficient, which is non-negligibly weighted with the atomic-scale variation in the coherent thermal transport through atom-to-atom thermal conductance [16] at the interface. Figure 2f shows a reconstructed image of theoretical thermoelectric voltage of $n$-doped free-standing graphene, simulated with $S^{sample}(\mathbf{r})$ and the linear fitting formula of $\Delta T_{coh}(\mathbf{r})$.

Using the linear correlation of $\Delta T_{coh}(\mathbf{r})$ with the vdW energy $E^{vdW}(\mathbf{r})$, we may be able to identify atomic-scale defects on graphene surface by comparing experimental thermoelectric images with simulated thermoelectric voltage images. Figure 3a and 3b show two independently scanned thermoelectric images of a point defect in bilayer graphene on SiC. The large-area and small-area scans produce almost identical thermoelectric images of the defect. The topographic analysis in Fig. 3c and 3d clearly indicates that the point defect in experiment is associated with a single carbon atom site. In order to map the defect with an atomic model, we simulated thermoelectric images of single carbon vacancy ($V_C$; Fig. 4a and 4b) [32], substitutional nitrogen ($N_C$; Fig. 4c and 4d) [33], and the defect complex [34] of a carbon vacancy and substitutional oxygen ($V_C$-$O_C$; Fig. 4e-4f) in $n$-doped free-standing graphene. Because the electronic state of $V_C$ locates below the Fermi energy (Fig. S3), its simulated images appear with a bright protrusion at the defect site, as opposed to the experimental image. On the other hand, the defect states of substitutional nitrogen $N_C$ and the defect complex $V_C$-$O_C$ locate above the Fermi energy (Figs. S4 and S5), which will cause a dark depression in the image because of the differential sensitivity [7]. Simulated thermoelectric images reproduce the negative depression of the experimental image at the defect site. While the size of the depressed region matches with $N_C$'s, the symmetry, localization, and electron scattering patterns of the thermoelectric image agree well with $V_C$-$O_C$'s. As atomic oxygen may be present during sample growth, the point defect is very likely to be $V_C$-$O_C$. The formation energy of self-passivated $V_C$-$O_C$ [34] is 3.0 eV, which is much smaller than the formation energy (7.6 eV) of $V_C$ [see Ref. 28]. With $V_{diff} = -0.6$ mV, we could reproduce some conspicuous features of the experimental image, as shown in Fig. S6.

Finally, it may be important to compare heat-based scanning thermoelectric microscopy with conventional STM. Both techniques share many common features and functionalities as types of scanning probe microscopy that exceptionally provide real-space images of wavefunctions. In an analogy, while STM measures tunneling current by applying a voltage drop across a vacuum-tunneling gap, thermoelectric microscopy measures voltage differences by applying a temperature drop across the interface or across a heat-transfer gap. As a result, the Fermi electrons are only perturbed at the first order by the temperature bias, in contrast to the zero-order perturbation by the voltage bias in STM. In this sense, scanning thermoelectric microscopy will be very useful for differentially analyzing the Fermi electronic states, even at room temperature.

**Acknowledgement**

We thank Y. Kuk, J. A. Stroscio, D. G. Cahill, C. K. Shih, J. Kang, M. Kang, and S. D. Kang for useful comments, and W. Kim for providing the sample. This work was supported by the NRF (2012R1A2A2A01046191) and Global Frontier R&D (2011-0031566) programs of MSIP. The work at KRISS was supported by the Converging Research Center Program (2013K000169) of MSIP.

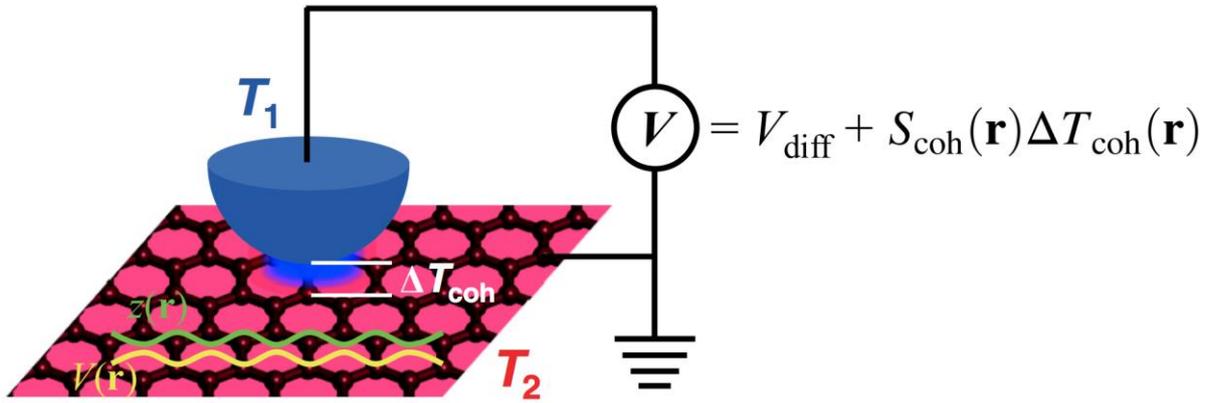

**Figure 1.** Schematic of the atomic-resolution scanning thermoelectric microscope. The graphene sample is at an elevated temperature $T_2$ in contact with the tip at temperature $T_1$. A temperature gradient field $\nabla T$ is developed in the vicinity of the tip-sample contact area, and more importantly across the interface between the tip and sample with an effective temperature drop $\Delta T_{coh}(\mathbf{r})$. The open-circuit thermoelectric voltage $V$ has a simple expression at the atomic scale as in Eq. (1). The measured $V(\mathbf{r})$ shows a phase difference of $180^o$ with atomic corrugation $z(\mathbf{r})$, as schematically shown in the figure.

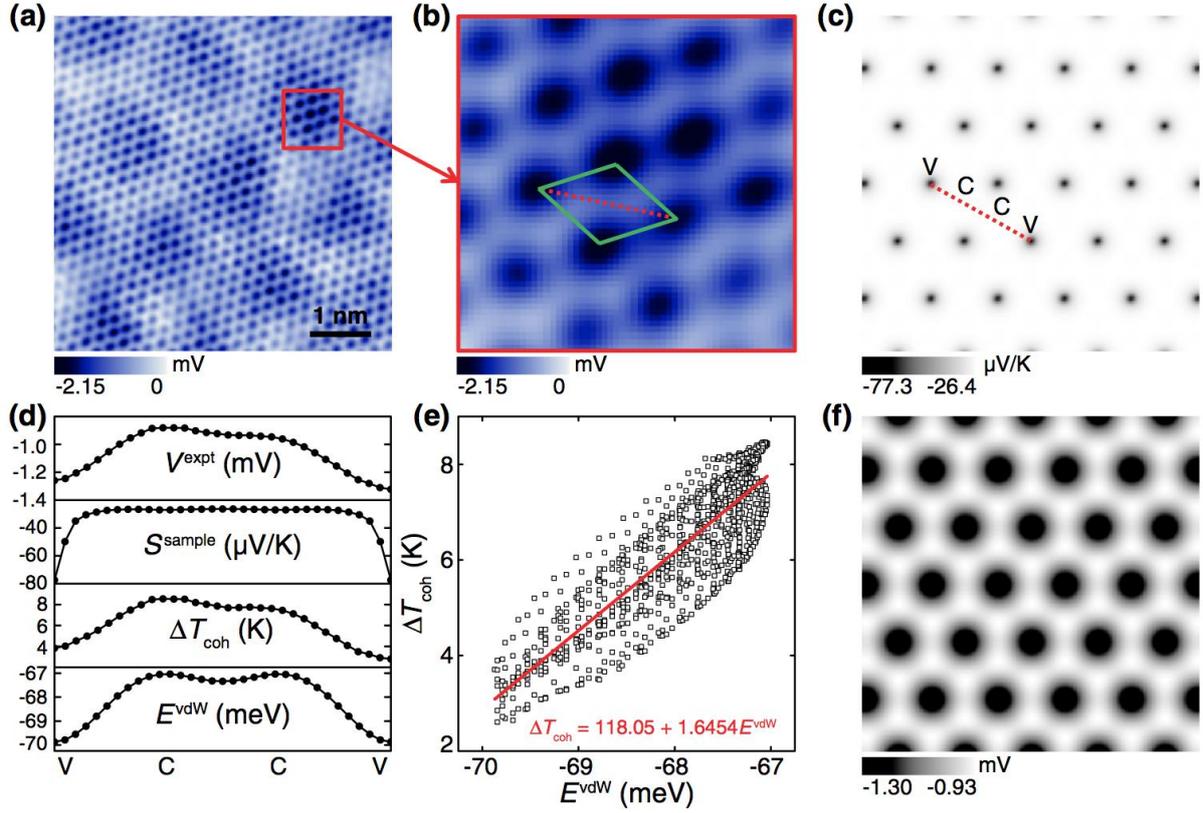

**Figure 2.** (a) Experimental thermoelectric voltage image for a defect-free region of bilayer graphene on SiC(0001), in which carbon hexagons are visible together with the interference pattern from the $6\sqrt{3}\times6\sqrt{3}R30°$ surface reconstruction of SiC substrate. (b) Enlarged area in (a). We sampled experimental thermoelectric voltages in a graphene unit cell, marked with a parallelogram. (c) Computer-simulated Seebeck coefficient image of $n$-doped free-standing graphene. The center void and the carbon atom are marked with 'V' and 'C', respectively. (d) Line profiles of experimental thermoelectric voltage, Seebeck coefficient, temperature drop, and van der Waals energy along the line of V–C–C–V in (c). (e) Correlation between the deduced $\Delta T_{coh}(\mathbf{r})$ and the van der Waals energy $E^{vdW}(\mathbf{r})$. Locally averaged $S^{sample}(\mathbf{r})$ with a disk radius of 0.5 Å was used for deducing $\Delta T_{coh}(\mathbf{r})$ (see Fig. S2). The linear fitting formula is $\Delta T_{coh}(\mathbf{r}) = 118.05 + 1.6454\,E^{vdW}(\mathbf{r})$. (f) Reconstructed image of theoretical thermoelectric voltage for $n$-doped free-standing graphene.

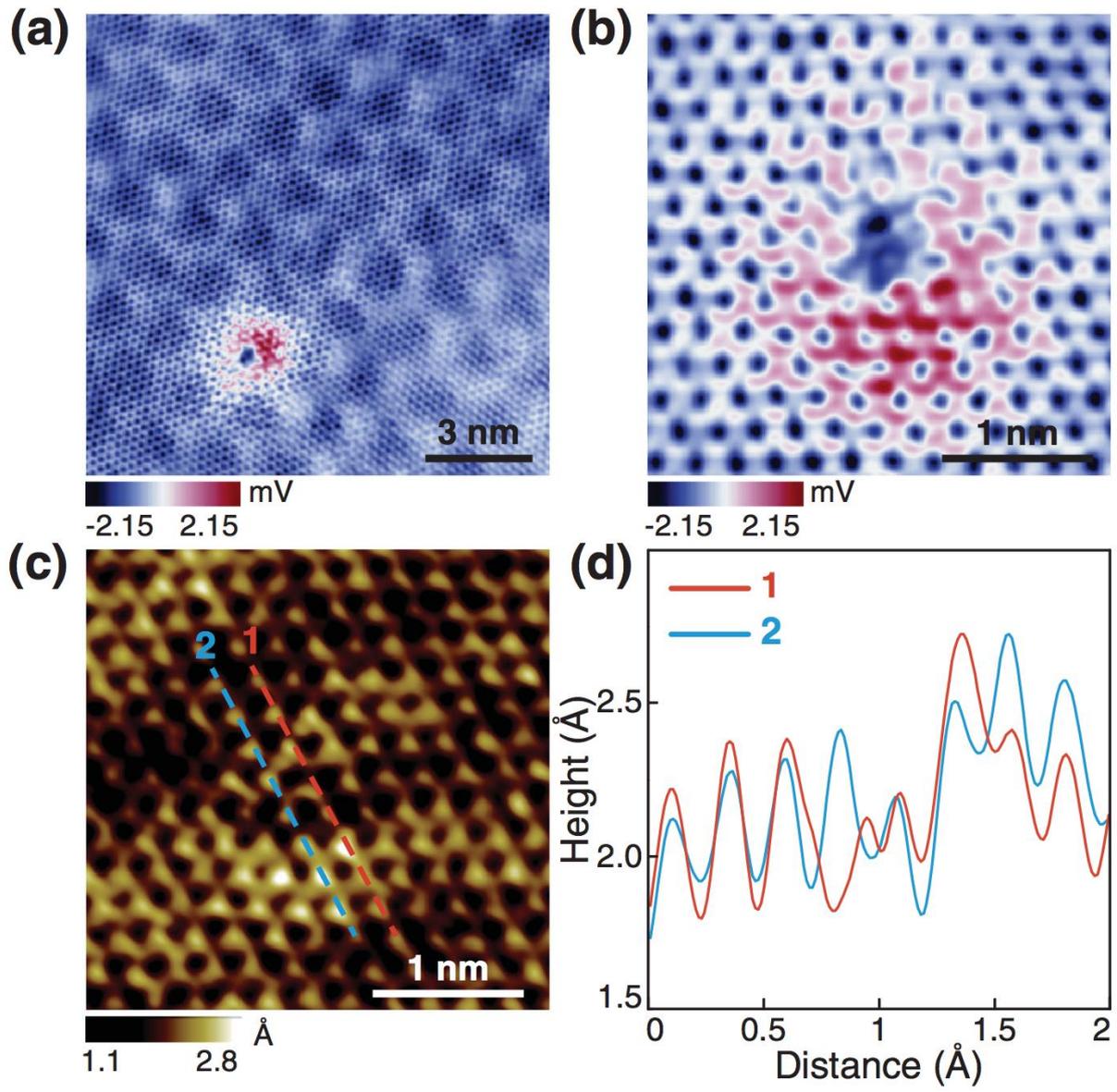

**Figure 3.** (a) Large-area scanning image of thermoelectric voltage for a point defect in bilayer graphene on SiC. (b) Small-area scanning image of thermoelectric voltage in a rotated view for the point defect in (a) and (c) simultaneously obtained topographic image. (d) Experimental height profiles along the red ('1') and blue ('2') dashed lines in (c). The profiling distance was measured from the top left along the lines shown in (c). One carbon atom site is clearly defective along the red line profile, as compared with the blue line profile of the neighboring atomic row.

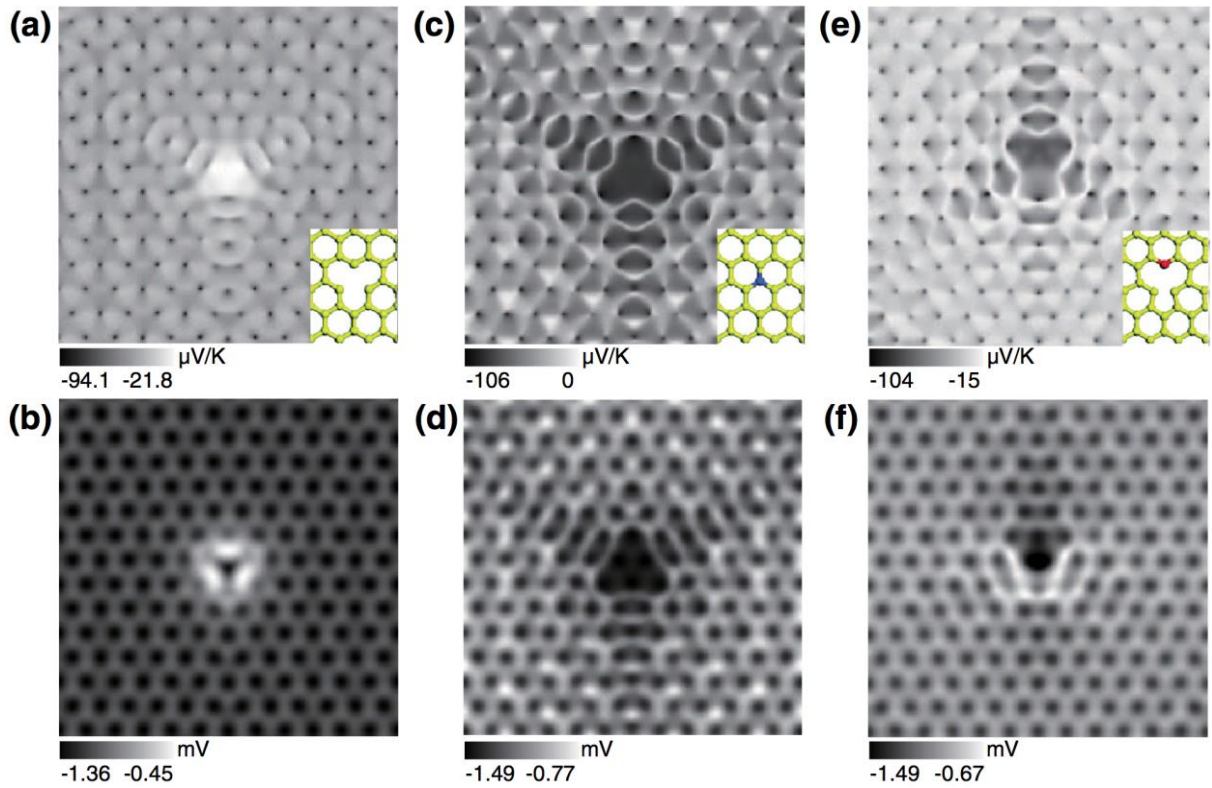

**Figure 4.** Simulated images of Seebeck coefficients (a,c,e) and thermoelectric voltages (b,d,f) near point defects of (a)-(b) a single carbon vacancy ($V_C$), (c)-(d) a substitutional nitrogen atom ($N_C$), and (e)-(f) a defect complex of a carbon vacancy and substitutional oxygen ($V_C$-$O_C$) in *n*-doped free-standing graphene. Atomic models are shown in the inset. See Fig. S6 for direct comparison with experiment.

## Supplemental Materials

# Seebeck Effect at the Atomic Scale

Eui-Sup Lee,[1] Sanghee Cho,[2] Ho-Ki Lyeo,[2,*] and Yong-Hyun Kim[1,*]

[1]*Graduate School of Nanoscience and Technology, KAIST, Daejeon 305-701, Korea*

[2]*Korea Research Institute of Standards and Science, Daejeon 305-340, Republic of Korea*

(Received: August 25, 2013)

*\*Correspondence and requests for materials should be addressed to*

Yong-Hyun Kim (yong.hyun.kim@kaist.ac.kr) or Ho-Ki Lyeo (hklyeo@kriss.re.kr)

**First-principles calculations.** The mesoscopic Seebeck coefficient in Eq. (6) was obtained using Kohn-Sham wavefunctions and local electronic density of states from the first-principles density-functional theory (DFT) calculations of graphene. We used projector-augmented wave (PAW) pseudopotentials and the Perdew-Burke-Ernzerhof (PBE) exchange-correlation functional for the ground-state total-energy calculations, as implemented in the VASP software [35]. We also employed a kinetic energy cutoff of 500 eV for plane wave expansions and (144×144×1) mesh equivalent **k**-points sampling per graphene unit cell for improved accuracy in the local density of states. We used the (12×12) graphene supercell for modeling the point defects ($V_C$, $N_C$, and $V_C$-$O_C$) in graphene. The defect formation energy was calculated from $E_{formation} = E(D) - \Sigma\, n_X \mu_X$, where $E(D)$ is the DFT total energy of the defect system, and $n_X$ and $\mu_X$ represent, respectively, the number and the chemical potential of element X. The chemical potentials of carbon and oxygen are taken from those of graphene and oxygen molecule. A Gaussian broadening of 0.05 eV and the Fermi energy $E_F$ = 0.3 eV above the Dirac point were used to calculate the local density of states for Eq. (6). We used $T$ = 315 K to calculate $S^{sample}(\mathbf{r})$ in Eq. (6).

**Effective temperature drops.** To deduce the effective temperature drop $\Delta T_{coh}(\mathbf{r})$ from experimental $V^{expt}(\mathbf{r})$ and theoretical $S^{sample}(\mathbf{r})$ by inverting Eq. (1), we have to use a locally-averaged $S^{sample}(\mathbf{r})$ within a certain disk radius $R$. The local averaging corresponds to the finite size effect of the voltage probe. In Fig. 2e-2f, we used locally-averaged $S^{sample}(\mathbf{r})$ of pristine graphene with $R$ = 0.5 Å. For comparison, we used the original $S^{sample}(\mathbf{r})$ and the locally-averaged $S^{sample}(\mathbf{r})$ with $R$ = 0.3 Å in Fig. S2 for constructing $\Delta T_{coh}(\mathbf{r})$ and reconstructing theoretical thermoelectric voltages.

The deduced $\Delta T_{\text{coh}}(\mathbf{r})$ shows an almost linear correlation with van der Waals (vdW) energy $E^{\text{vdW}}(\mathbf{r})$, which was calculated with Lennard-Jones 12-6 potentials;

$$E^{\text{vdW}}(\mathbf{r}_i) = \sum_j 4\varepsilon_{ij}\left[\left(\frac{\sigma_{ij}}{|\mathbf{r}_i - \mathbf{r}_j|}\right)^{12} - \left(\frac{\sigma_{ij}}{|\mathbf{r}_i - \mathbf{r}_j|}\right)^{6}\right], \tag{7}$$

where $\mathbf{r}_i$ and $\mathbf{r}_j$ are the atomic positions of the tip and the sample, respectively, and $\varepsilon$ and $\sigma$ are the Lennard-Jones parameters. The tip was modelled as single carbon atom. The $\varepsilon_{ii}$ and $\sigma_{ii}$ parameters for carbon, nitrogen, and oxygen atoms are listed in Table S1, and we used $\varepsilon_{ij} = \sqrt{(\varepsilon_{ii}\,\varepsilon_{jj})}$ and $\sigma_{ij} = (\sigma_{ii} + \sigma_{jj})/2$. The vdW energy was summed when the atom-atom distance is less than 15 Å, and the equilibrium height $z(\mathbf{r})$ at the minimum vdW energy $E^{\text{vdW}}(\mathbf{r})$ was used for evaluating $S^{\text{sample}}(\mathbf{r})$.

**Table S1.** Lennard-Jones parameters for C, N, and O atoms [29,30].

| $i$ | $\varepsilon_{ii}$ (meV) | $\sigma_{ii}$ (Å) |
|---|---|---|
| C | 4.20 | 3.37 |
| N | 7.41 | 3.25 |
| O | 9.12 | 2.96 |

**Seebeck coefficient and Fermi temperature.** We further discuss the physics of the sample Seebeck coefficient, expressed as in Eq. (6), or

$$S^{\text{sample}}(\mathbf{r}) = -\frac{(k_B\pi)^2 T}{3e}\frac{\partial}{\partial E}\left(\ln\left[N_e^{\text{sample}}(E,\mathbf{r})\right]\right)_{E_F}. \tag{8}$$

If we define $k_B \ln[N_e^{\text{sample}}(E,\mathbf{r})]$ as the electronic entropy, the energy derivative of the entropy could correspond to a reciprocal temperature, expressed as

$$\frac{1}{T_F(\mathbf{r})} = \frac{\partial}{\partial E}\left(k_B \ln\left[N_e^{\text{sample}}(E,\mathbf{r})\right]\right)_{E_F}, \tag{9}$$

where $T_F(\mathbf{r})$ is the position-dependent statistically-defined Fermi temperature of a material. The newly defined Fermi temperature $T_F(\mathbf{r})$ of a real material is a generalized version of the 'Fermi temperature' $T_F = E_F/k_B$ that was only applicable for a three-dimensional free electron gas model [36]. Then, the sample Seebeck coefficient can be generally written as,

$$S^{\text{sample}}(\mathbf{r}) = -\frac{\pi^2}{3}\frac{k_B}{e}\frac{T}{T_F(\mathbf{r})}, \tag{10}$$

which can be interpreted as the ratio of the thermal equilibrium temperature ($T$) and Fermi temperature ($T_F$) of a material. Note that the statistically-defined Fermi temperature is a material property, not a real temperature, which can be either positive or negative depending on the slope of $N_e^{sample}(E,\mathbf{r})$. The Seebeck coefficient in Eq. (10) may be conceptually connected to other electron-related thermal characteristics of a material, such as the electronic heat capacity $c_v = \pi^2 N_e^{sample}(E_F) k_B^2 T / 3$ [36] and the quantum of thermal conductance $g_0 = \pi^2 k_B^2 T / 3h$ [37]. For example, the Seebeck coefficient and electronic heat capacity can be linked as

$$S^{sample}(\mathbf{r}) = \frac{c_v}{e} \frac{\partial}{\partial E} \left( \frac{1}{N_e^{sample}(E,\mathbf{r})} \right)_{E_F}. \tag{11}$$

**Supplemental Figures:**

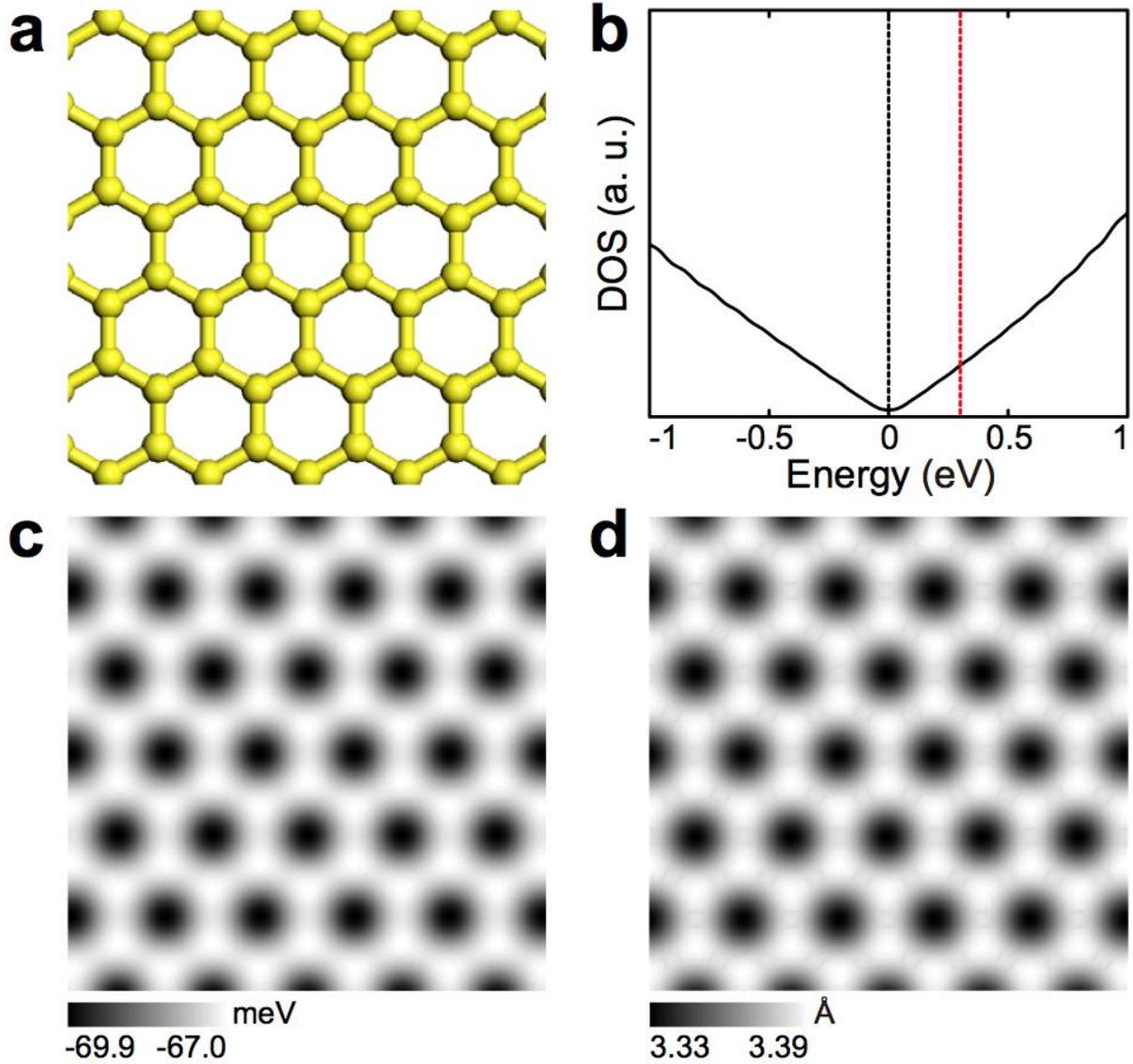

**Figure S1.** (a) Ball-and-stick model and (b) density of states (DOS) of pristine free-standing graphene. The zero energy indicates the charge-neutrality point or the Dirac point (the black dashed line). The red dashed line marks the Fermi energy (0.3 eV) used for the thermoelectric simulation in Fig. 2. Computer-simulated images of (c) van der Waals energy $E^{vdW}(\mathbf{r})$ and (d) van der Waals topography $z(\mathbf{r})$ at the minimum energy.

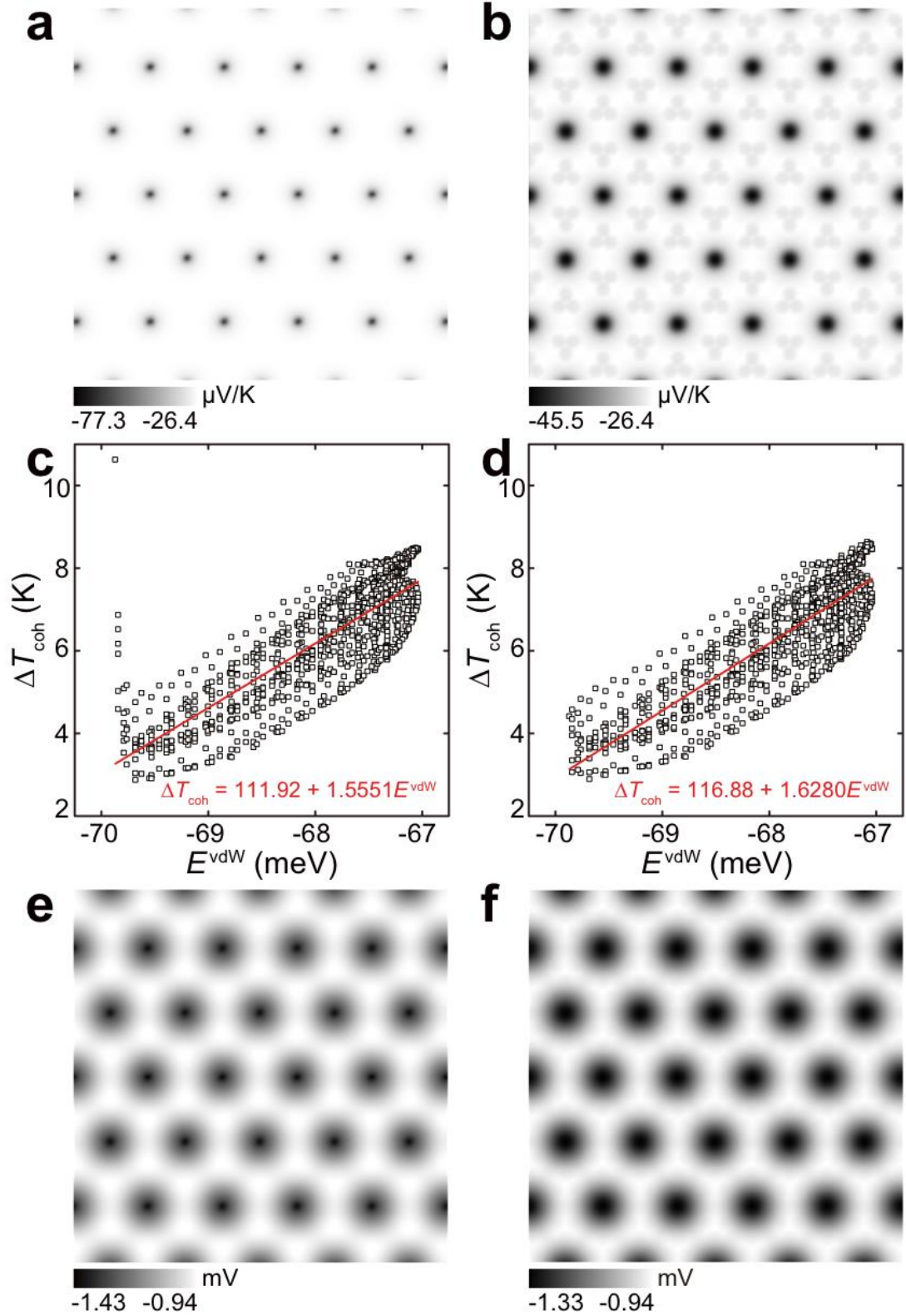

**Figure S2.** Computer-simulated images of locally-averaged Seebeck coefficients within a disk radius (a) $R = 0$ Å (or no average) and (b) $R = 0.3$ Å. Correlation between van der Waals energy and effective temperature drop deduced with (c) $R = 0$ Å and (d) $R = 0.3$ Å. Reconstructed images of thermoelectric voltages for (e) $R = 0$ Å and (f) $R = 0.3$ Å.

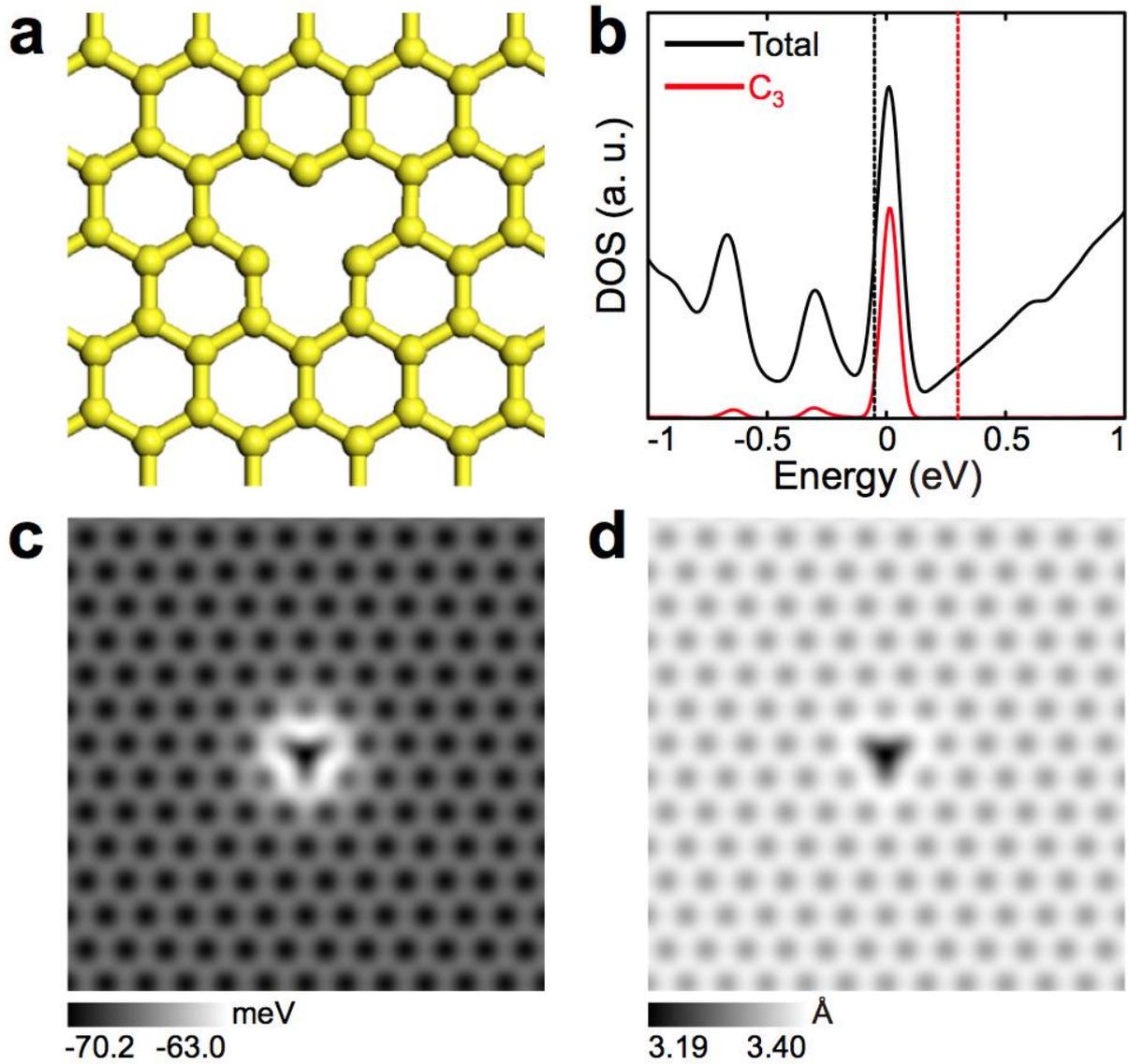

**Figure S3.** (a) Ball-and-stick model and (b) density of states (DOS) of defective free-standing graphene with a single carbon vacancy ($V_C$). In (b), the local DOS of the dangling-bonded $C_3$ atoms is displayed. The zero energy indicates the original Dirac point; the black dashed line indicates the original Fermi energy of the defective graphene; the red dashed line indicates the elevated Fermi energy (0.3 eV) used for the thermoelectric simulation (Fig. 4a-4b). Computer-simulated images of (c) van der Waals energy $E^{vdW}(\mathbf{r})$ and (d) van der Waals topography $z(\mathbf{r})$ at the minimum energy.

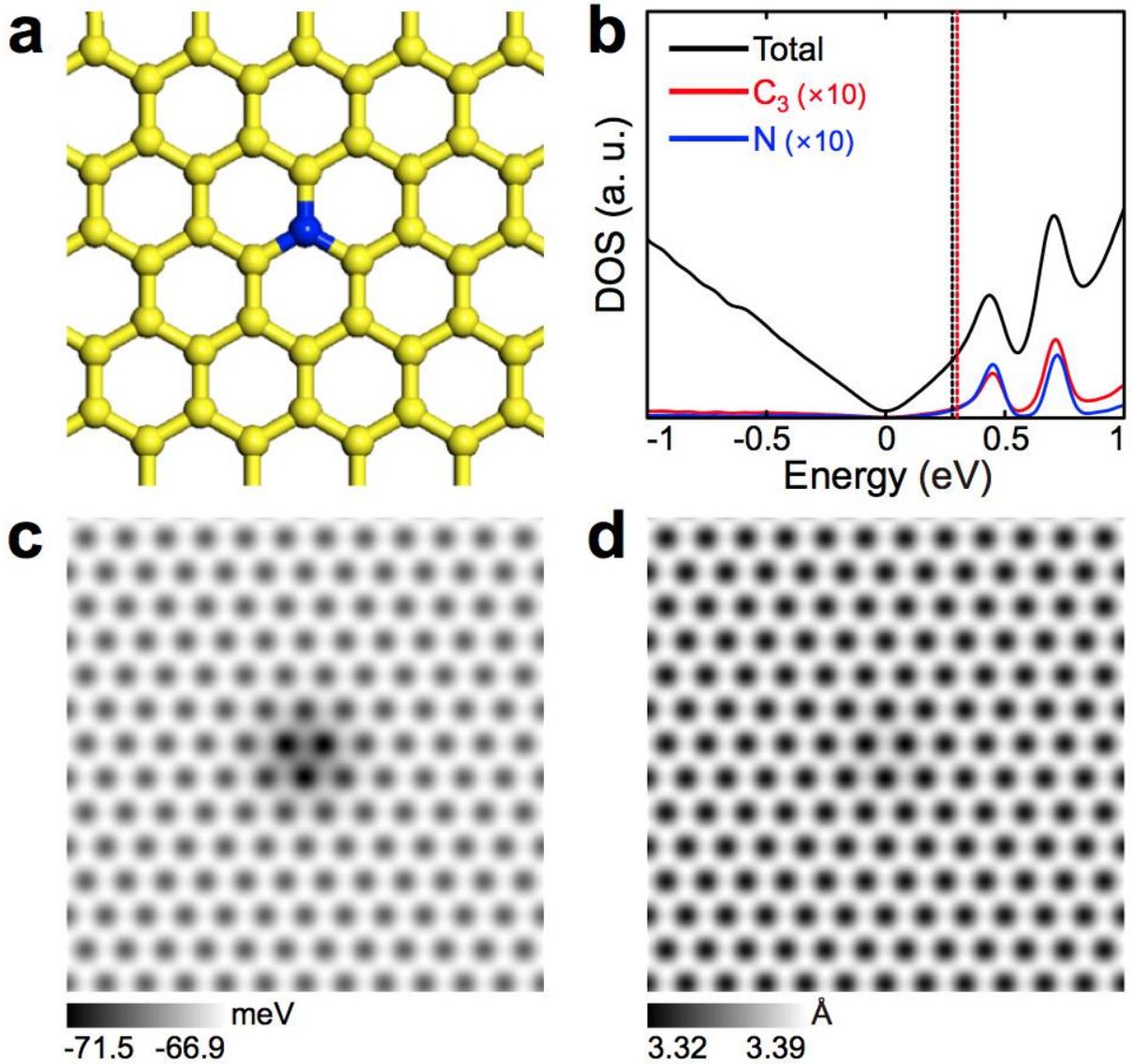

**Figure S4.** (a) Ball-and-stick model and (b) density of states (DOS) of defective free-standing graphene with a single substitutional nitrogen atom ($N_C$), marked in blue. In (b), the local DOSs for the nitrogen atom and the three neighboring carbon atoms ($C_3$) are displayed after being multiplied by 10 for clarity. The zero energy indicates the original Dirac point; the black dashed line indicates the original Fermi energy of the defective graphene; the red dashed line indicates the Fermi energy (0.3 eV) used for the thermoelectric simulation (Fig. 4c-4d). Computer-simulated images of (c) van der Waals energy $E^{vdW}(\mathbf{r})$ and (d) van der Waals topography $z(\mathbf{r})$ at the minimum energy.

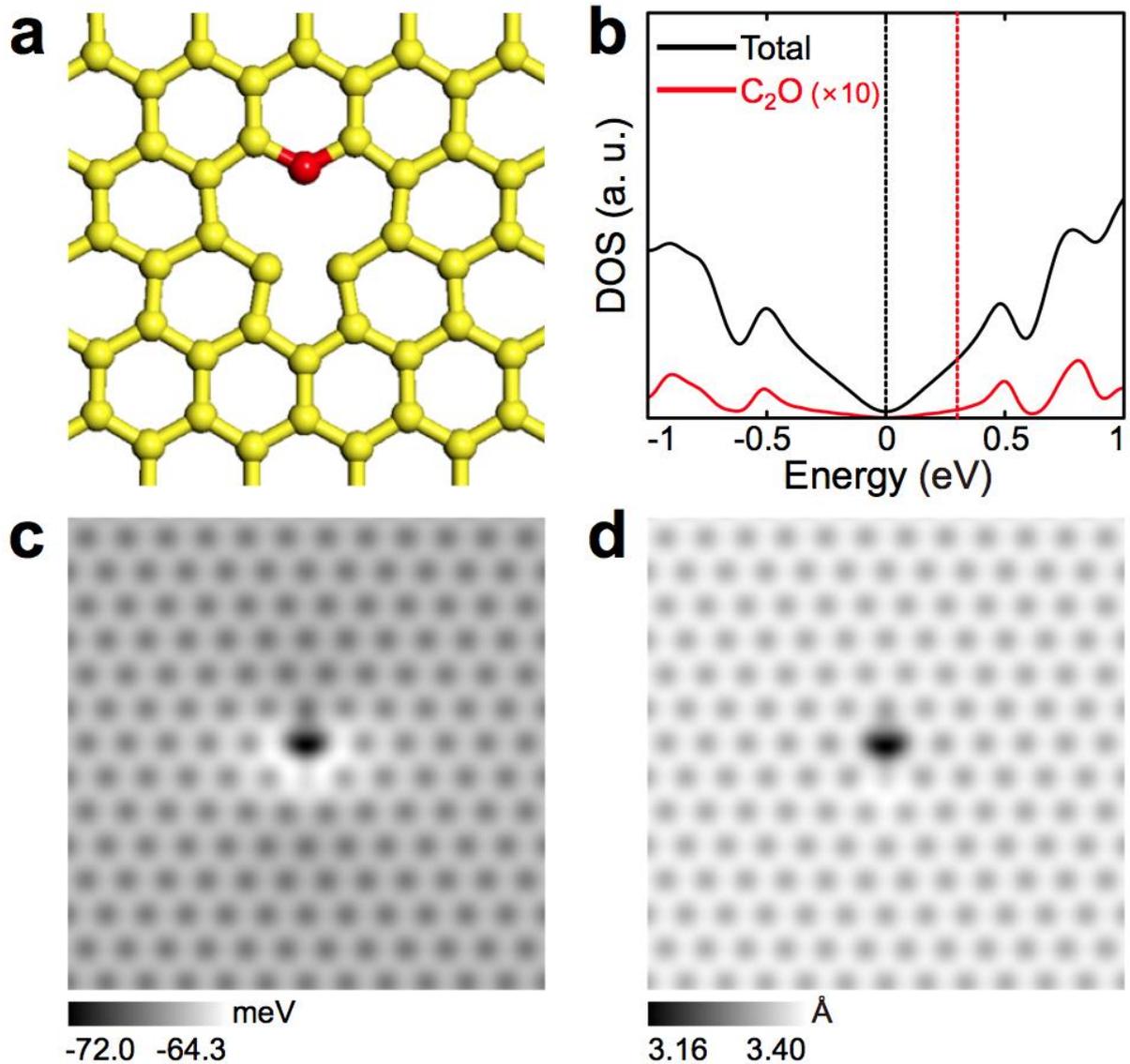

**Figure S5.** (a) Ball-and-stick model and (b) density of states (DOS) of defective free-standing graphene with the defect complex $V_C$-$O_C$, in which oxygen atom is marked in red. The defect complex is mostly self-passivated [34]. In (b), the local DOS for the $C_2O$ atoms around the vacancy is displayed after being multiplied by 10 for clarity. The zero energy indicates the original Dirac point; the black dashed line indicates the original Fermi energy of the defective graphene; the red dashed line indicates the Fermi energy (0.3 eV) used for the thermoelectric simulation (Fig. 4e-4f). Computer-simulated images of (c) van der Waals energy $E^{vdW}(\mathbf{r})$ and (d) van der Waals topography $z(\mathbf{r})$ at the minimum energy.

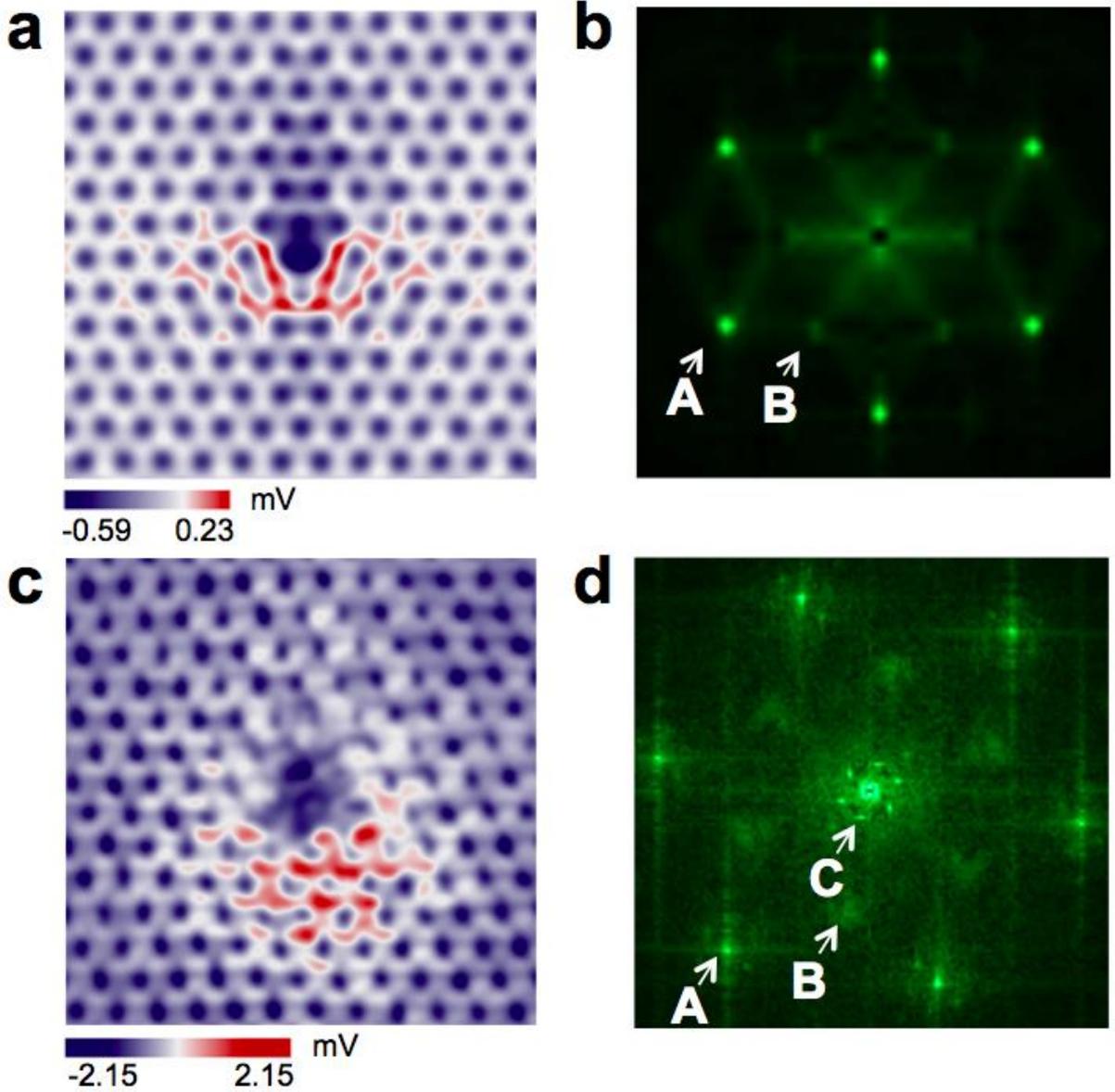

**Figure S6.** (a) Computer-simulated image of thermoelectric voltages for the defect complex $V_C$-$O_C$ with $V_{diff} = -0.6$ mV, and (b) its Fast Fourier Transform (FFT) image. In order to compare with the experimental image, we generated a theoretical image of positively shifted thermoelectric voltage in (a). In (b), the arrow A corresponds to the peak from the reciprocal lattice, and six B peaks appear at the corners of the Brillouin zone due to the Fermi wave vector $k_F$ intervalley scattering [38]. (c) Experimental thermoelectric image of Fig. 3b, color-rescaled here for convenience, and (d) the FFT image of Fig. 3a. In (d), the arrow A corresponds to the peak from the reciprocal lattice. The arrow B approximately corresponds to the Fermi wave vector $k_F$, and the six B peaks correspond to the corners of the Brillouin zone. The arrow C represents the 6×6 pattern with respect to SiC [39], originated from the $6\sqrt{3} \times 6\sqrt{3}R30°$ reconstruction of the SiC(0001) surface. Note that this pattern is oriented

identically to the peaks from the Brillouin zone. The positive thermoelectric signals near the defect site in experiment may associate with fine details of wavefunction overlap, atom-to-atom thermal coupling, substrate effect, and diffusive shift $V_{\text{diff}}$, which may be limitedly reflected in our current simulation scheme.

**Supplemental References:**